\documentclass{cimento}

\usepackage{graphicx}  % got figures? uncomment this
\title{Measurements in charmless $B$ decays at Belle~II}
\author{R.~Manfredi\thanks{On behalf of the Belle~II collaboration.}}
\instlist{
  \inst{} University and INFN Trieste - Trieste, Italy}
\begin{document}

\maketitle

\begin{abstract}
We report on preliminary measurements of branching fractions, charge-parity-violating asymmetries, and longitudinal polarization fractions in charmless bottom-meson decays from the Belle~II experiment. We use samples of electron-positron collisions collected in 2019 and 2020 at the $\Upsilon(4S)$ resonance,  corresponding to integrated luminosities of up to 62.8 ${\rm fb^{-1}}$. The results are compatible with known values, indicating good understanding of early detector performance.
\end{abstract}

\section{Introduction}
The physics of charmless bottom-meson ($B$) decays plays an important role in the Belle~II physics program, with results expected to improve the precision of several sensitive tests of the Standard Model in the flavor sector. 
Belle~II goals in charmless $B$ physics exploit the unique experimental features of low-background and precise knowledge of the center-of-mass energy, and include tests of isospin sum-rules,  investigations of localized charge-parity (CP) violating asymmetries in the phase space of multibody $B$ decays,  and improved  determinations of the CKM angle $\alpha$/$\phi_2 \equiv \arg\left[-V_{td}V^*_{tb}/V_{ud}V^*_{ub}\right]$~\cite{kou}.\\
Belle~II is a particle detector designed to study 7-on-4 GeV electron-positron collisions at 10.58 GeV,  produced at very high luminosity by the SuperKEKB collider located at the KEK laboratory in Japan~\cite{akai}.  The collision energy corresponds to the $\Upsilon(4S)$ resonance mass, which decays almost exclusively to $B\bar{B}$ pairs with little available energy to produce additional particles, resulting in low backgrounds. The beam-energy asymmetry boosts the center of mass allowing displacements of the $B$ decay vertexes, for decay-time dependent measurements. \\
Belle II consists of several subdetectors, arranged hermetically in a cylindrical geometry around the interaction point. The innermost detector is a silicon tracker, based on pixel sensors for the first two layers and on silicon strip sensors for the surrounding four layers.  The vertex detector samples the trajectories of final-state charged particles at radii $1.4<r<13.5~{\rm cm}$ to reconstruct the decay position (vertex) of their long-lived parent particles, with a resolution of about 15 $\rm \mu m$. A large-radius wire drift chamber measures charged-particle charges, momenta with 0.4\% resolution, and $dE/dx$ with 5\% resolution. A time-of-propagation Cherenkov detector and an aerogel ring-imaging Cherenkov detector surround the drift chamber and provide charged-particle identification information, allowing for separation of kaons from pions of up to 4 GeV/$c$ momentum, with 90\% efficiency and 5\% misidentification rate.  A CsI(Tl)-crystal electromagnetic calorimeter measures the energy of electrons and photons, with 1.6\%--4\% resolution. Layers of plastic scintillators and resistive-plate chambers alternated with iron plates provide muon and $K_{\rm L}^0$ reconstruction.\\
Physics operations started on March 11,  2019, aiming at collecting a sample comparable in size to the ones from previous $B$ factories by summer 2022, and with the goal of collecting fifty times more data by 2031.  
The data used in this work have been collected up to May 14, 2020 (``summer data set"), corresponding to an integrated luminosity of 34.6 ${\rm fb^{-1}}$,  or up to July 1, 2020 (``full data set"), corresponding to 62.8 ${\rm fb^{-1}}$~\cite{lumi}.\\
We reconstruct decays with branching fractions of $10^{-6}$ or larger,  with expected yields sufficient to obtain visible signals in the available data. We study $B^0\to K^+\pi^-$,  $B^+\to K^+\pi^0$, $B^+\to K^{0}_{\rm S}\pi^+$, and $B^0\to K^{0}_{\rm S}\pi^0$ decays targeting the $K\pi$ isospin sum-rule~\cite{gronau}; $B^+\to K^+K^-K^+$ and $B^+\to K^+\pi^-\pi^+$ decays as a first step toward investigations of local CP violation; and $B^0\to \pi^+\pi^-$, $B^+\to \pi^+\pi^0$, and $B^+\to \rho^+\rho^0$ decays in preparation for the determination of $\alpha$/$\phi_2$. Charge-conjugate modes are implied, except when noted otherwise.\\

\section{Analysis strategy}
The workflow is similar for all reported measurements. 
The main challenge is to improve the initial $10^{-5}$ signal-to-background ratios with sufficiently discriminating selections that isolate abundant, low-background signals. \\
We form final-state particle candidates by applying baseline selection criteria, then combine candidates in kinematic fits consistent with the topology of the desired decays to reconstruct $B$ candidates. 
The dominant background ({\it continuum}) comes from random combinations of particles produced in \mbox{$e^+e^-\to q\bar{q}$} \mbox{$(q=u,d,s,c)$} processes, which have a four times larger cross-section than $e^+e^-\to \Upsilon(4S)$.  We use a binary boosted decision-tree classifier to suppress continuum, which makes a non-linear combination of approximately 40 discriminating variables based on kinematics, decay time or event shape. We train the classifier to identify signal and continuum features using simulated samples. 
Additional background ($B\bar{B}$) comes from combinations of final-state particles of non-signal $B$ decays in $e^+e^-\to \Upsilon(4S)\to B\bar{B}$ processes. 
We vary the selection simultaneously on continuum-suppression output and charged-particle identification information to maximize $S$/$\sqrt{S+B}$, were $S$ and $B$ are signal and background yields, respectively, expected from simulation. The $\pi^0$ selection is optimized using \mbox{$B^+\to\bar{D}^0(\to K^+\pi^-\pi^0)\pi^+$} decays in data.\\
More than one candidate per event may be reconstructed. In the channels with a $\pi^0$ in the final state, one $\pi^0$ candidate per event is chosen based on the best kinematic fit. In all channels, one $B$ candidate per event is chosen randomly after applying the optimized selection, except for the $B^+\to\rho^+\rho^0$ channel where the selection is based on the best vertex fit. \\
After selection, signal yields are determined by fitting the beam-energy-constrained mass, $M_{\rm bc} \equiv \sqrt{s/(4c^4) - (p^{*}_B/c)^2}$, and energy difference, $\Delta E \equiv E^{*}_{B} - \sqrt{s}/2$, where $\sqrt{s}$ is the collision energy, and $E^{*}_{B}$ and $p^{*}_{B}$ are the reconstructed energy and magnitude of momentum of $B$ meson candidates, respectively, all in the $\Upsilon(4S)$ frame. $M_{\rm bc}$ separates continuum from $B\bar{B}$ events, while $\Delta E$ adds further discrimination from other background $B$ decays. 
Wrongly reconstructed signal decays ({\it self cross-feed}) are included in the fit, with constraints based on simulation expectations. \\
We determine the branching fractions $\mathcal{B}=N/\left(2\varepsilon N_{B\bar{B}}\right)$, where $N$ is the signal yield obtained from the fit; $\varepsilon$ is the reconstruction and selection efficiency, determined from simulation to be typically 10--30\%;  and $N_{B\bar{B}}$ is the number of $B\bar{B}$ pairs, 35.8  million  for $B^+B^-$ and 33.9 million for $B^0\bar{B}^0$ pairs.  The values of $N_{B\bar{B}}$ are determined from the measured integrated luminosity, the \mbox{$e^+e^-\to\Upsilon(4{\rm S})$}~cross section~$(1.110 \pm 0.008)\,$nb~\cite{Bevan:2014iga}~(assuming that the $\Upsilon(4{\rm S})$ decays exclusively to $B\bar{B}$~pairs), and the \mbox{$\Upsilon(4{\rm S})\to B^0\bar{B}^0$} branching fraction $f^{00} = 0.487\pm 0.010\pm 0.008$~\cite{Aubert:2005bq}.
We also measure CP-violating asymmetries $\mathcal{A}_{\rm CP} = \mathcal{A} - \mathcal{A}_{\rm det}$, where $\mathcal{A}$ is the flavor-specific signal yield asymmetry, and \mbox{$\mathcal{A}_{\rm det}\approx$ 0.7}--1.5\% is the instrumental asymmetry due to differences in interaction or reconstruction probabilities between particles and antiparticles. We measure $\mathcal{A}$ by determining simultaneously the flavor-specific signal yields, using the asymmetry as a fit parameter. We determine the instrumental asymmetries by measuring the flavor asymmetries in abundant samples of $D$ decays, assuming no CP violation. 
We finally assess the main systematic uncertainties based on experiments using data generated with a simplified simulation or studies on control modes.  The uncertainties are all dominated by sample size.

\section{Towards the $K\pi$ isospin sum-rule}
\begin{figure}[h!]
\centering
\includegraphics[width=5.75cm]{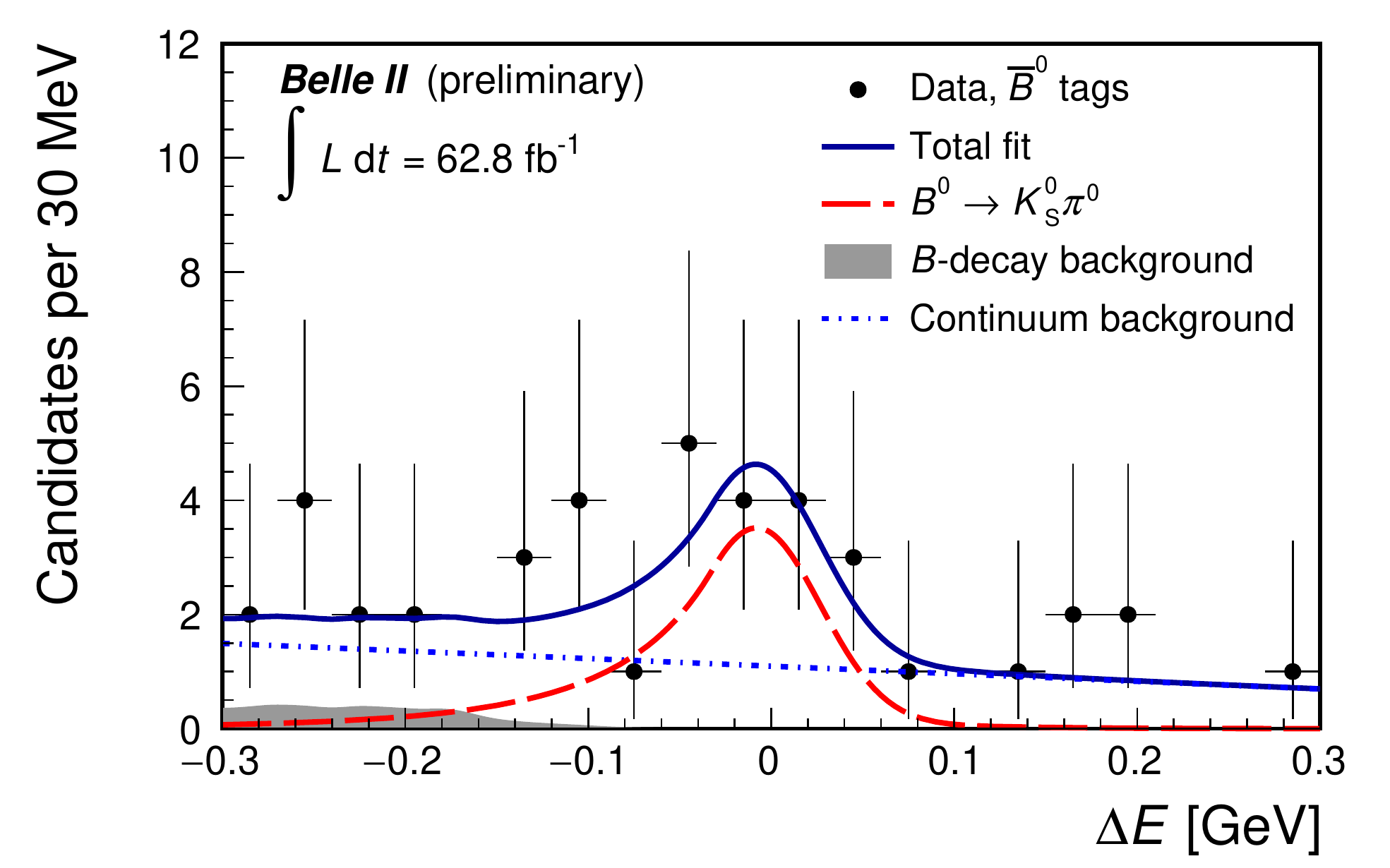} 
 \includegraphics[width=5.75cm]{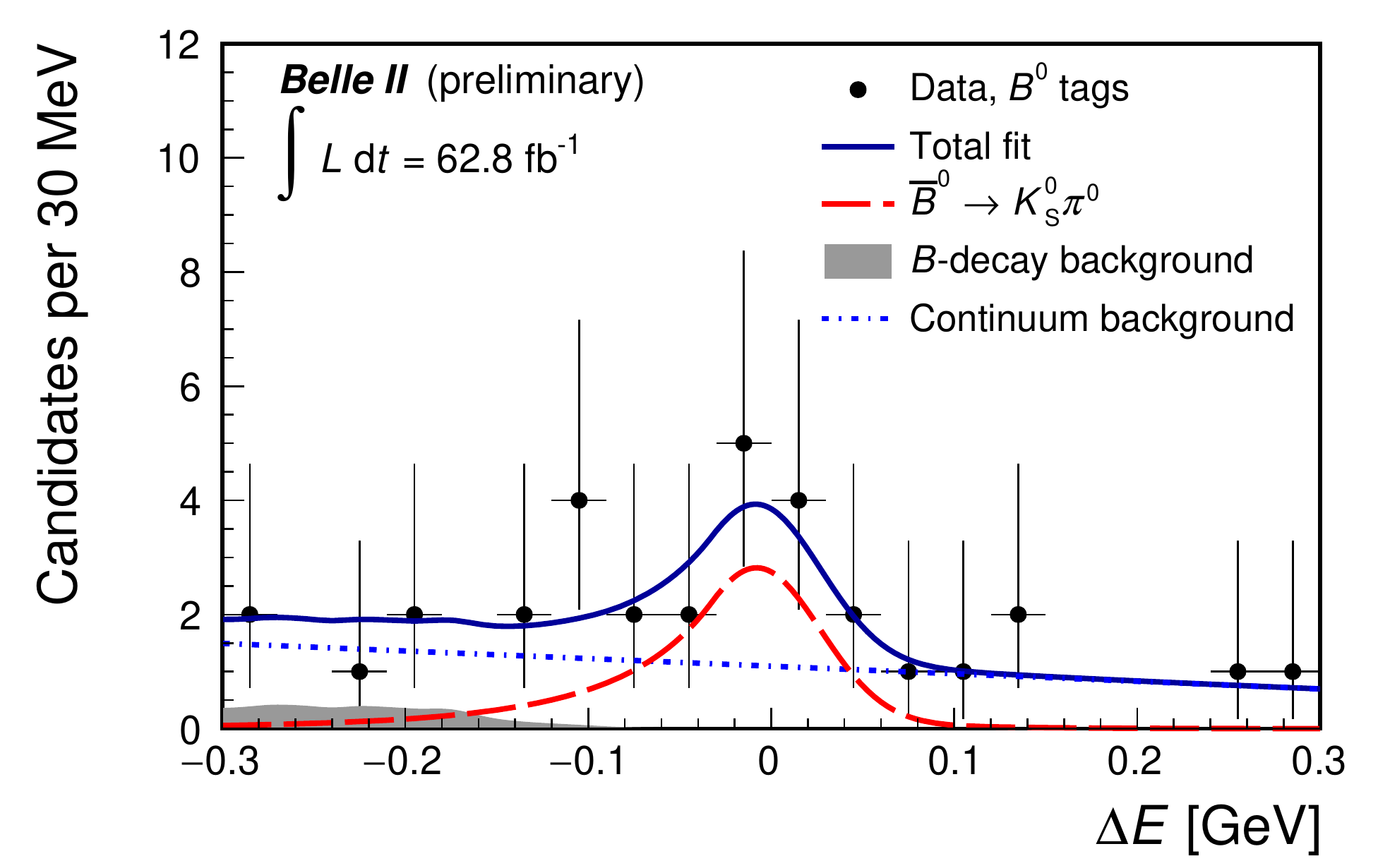}
 \includegraphics[width=5.75cm]{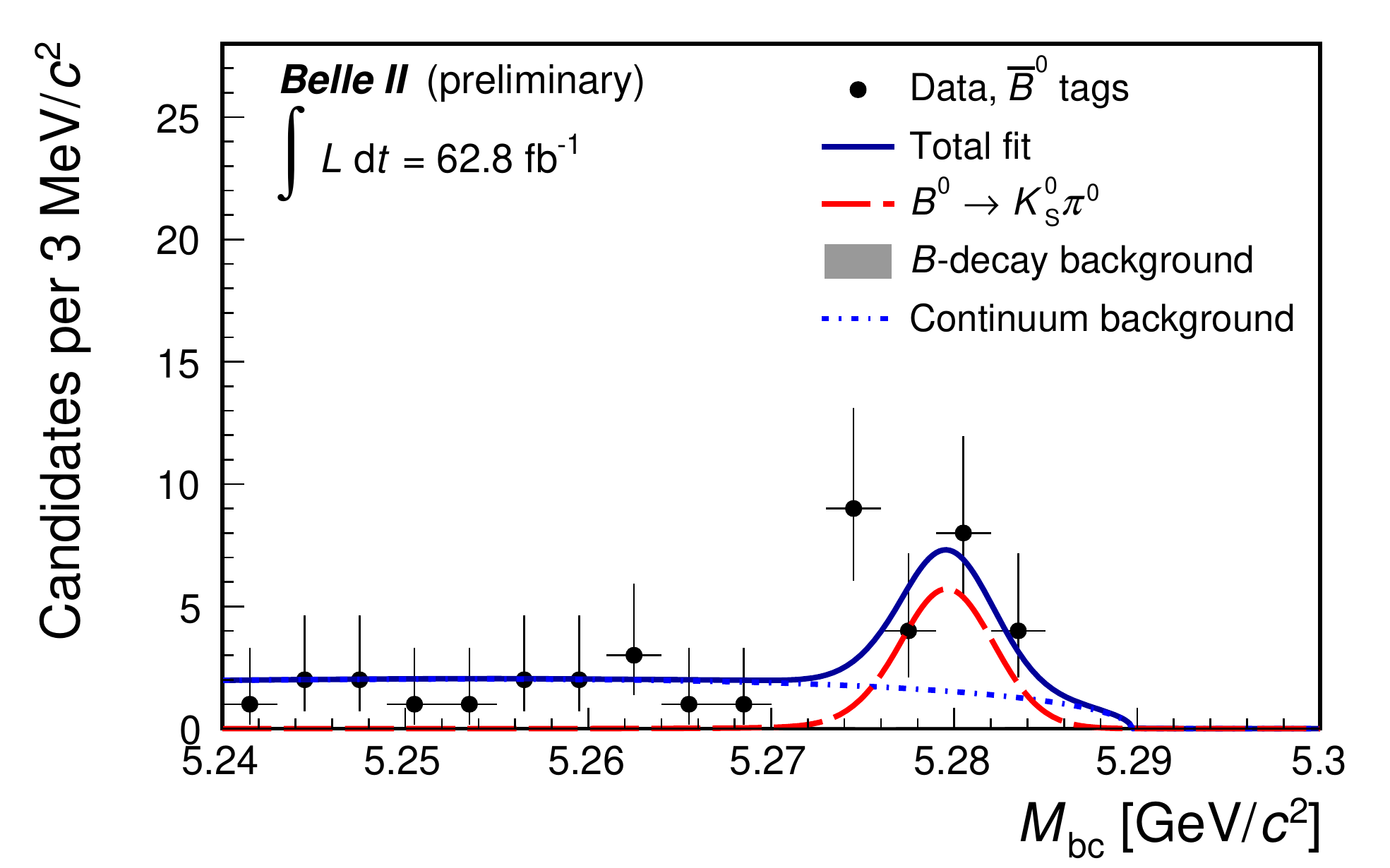} 
 \includegraphics[width=5.75cm]{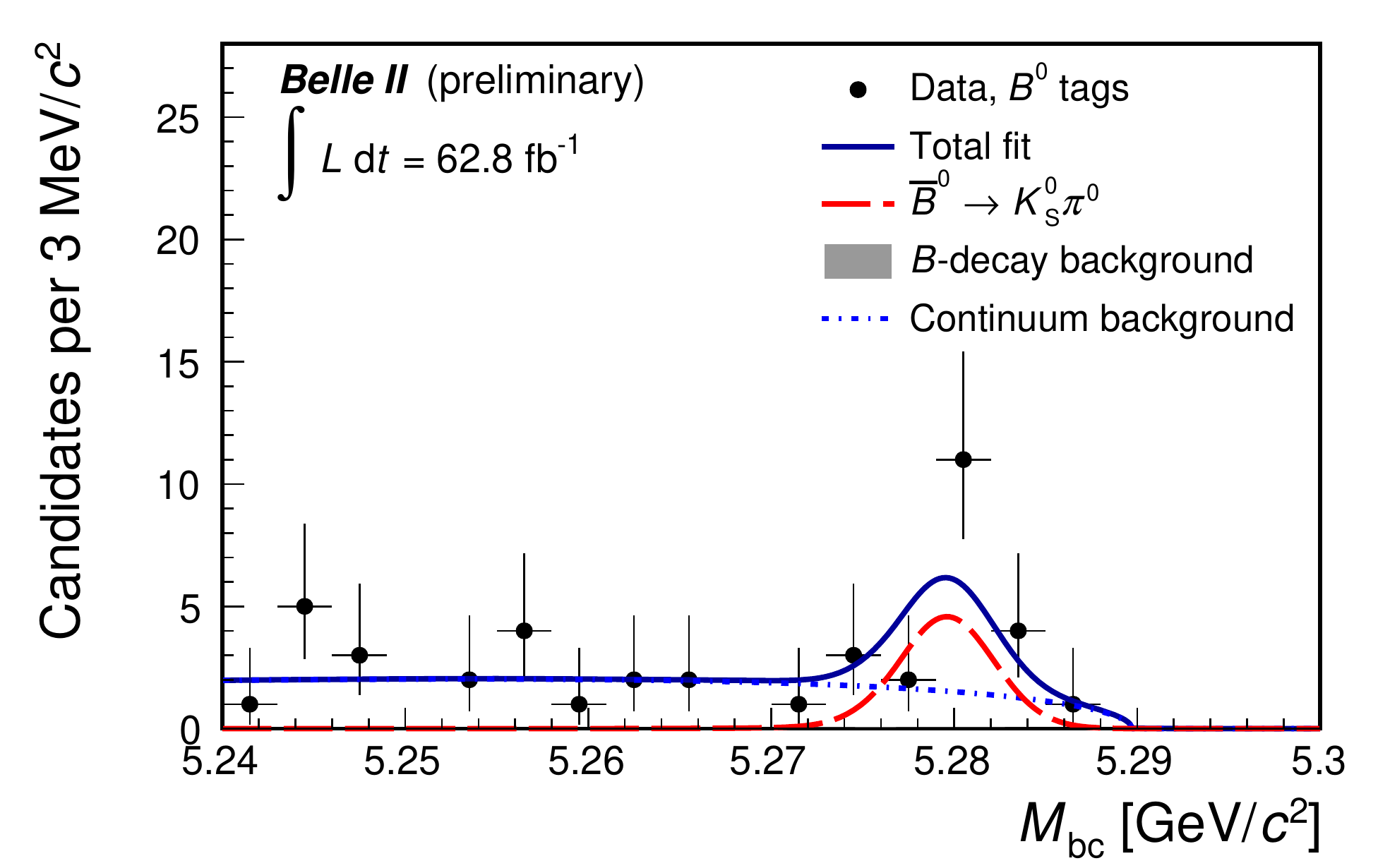} 
 \caption{Signal-enhanced distributions of (top) $\Delta E$ and (bottom) $M_{\rm bc}$ for (left) $B^0\to K^0\pi^0$ and (right) \mbox{$\bar{B}^0\to \bar{K}^0\pi^0$} candidates reconstructed in the 2019--2020 Belle~II data. The reconstructed flavor of the partner bottom meson in the event, used to separate the flavor-specific distributions, is indicated in the legends as ``tags". Fit projections are overlaid.}
\label{fig:K0pi0}
\end{figure}

Isospin sum-rules are reliable dynamical relations that combine branching fractions and CP-violating asymmetries of isospin-partner decays, properly accounting for subleading amplitudes.  The $K\pi$ sum-rule, $$I_{K\pi}=A_{\rm CP}^{K^+\pi^-} +A_{\rm CP}^{K^0\pi^+}\frac{\mathcal{B}(K^0\pi^+)}{\mathcal{B}(K^+\pi^-)}\frac{\tau_{B^0}}{\tau_{B^+}}-2A_{\rm CP}^{K^+\pi^0}\frac{\mathcal{B}(K^+\pi^0)}{\mathcal{B}(K^+\pi^-)}\frac{\tau_{B^0}}{\tau_{B^+}} -2A_{\rm CP}^{K^0\pi^0}\frac{\mathcal{B}(K^0\pi^0)}{\mathcal{B}(K^+\pi^-)},$$ 
was proposed to address the anomalously large difference observed between $\mathcal{A}_{\rm CP}(B^0\to K^+\pi^-)$ and $\mathcal{A}_{\rm CP}(B^+\to K^+\pi^0)$ values~\cite{gronau}, and offers  a stringent null test of the SM. \\
Results on $B^0\to K^+\pi^-$, $B^+\to K^+\pi^0$, and $B^+\to K^{0}_{\rm S}\pi^+$ decays are obtained on the summer data set~\cite{charmless1}, while measurements of $B^0\to K^{0}_{\rm S}\pi^0$ are performed with the full data set. 
For the summer data set results, we determine signal yields from maximum likelihood fits of the unbinned $\Delta E$ distribution, restricted in the $M_{\rm bc}$ signal region. For the $B^0\to K^{0}_{\rm S}\pi^0$ measurements, we include also $M_{\rm bc}$ in a two-dimensional fit (Fig. \ref{fig:K0pi0}).\\
The final state $K^{0}_{\rm S}\pi^0$ is a CP eigenstate. $B^0$ and $\bar{B}^0$ decays are therefore separated using flavor tagging information~\cite{tagger}. 
In the fit performed to determine $\mathcal{A}_{\rm CP}$, the $\Delta E$-$M_{\rm bc}$  probability density function is multiplied by the flavor tagger output, $$P_{\rm sig}(q)=\frac{1}{2}\left(1+q\cdot(1-2w_r)\cdot(1-2\chi_d)\cdot\mathcal{A}_{\rm CP}(K^0\pi^0)\right),$$ where $q=\pm1$ is the flavor tagger output determined on data,  $w_r\approx$ 0.00--0.47 is the fraction of wrong-tagged events determined on control modes~\cite{tagValidation}, $\chi_d=0.1858\pm0.0011$ is the known time-integrated $B^0$-mixing probability~\cite{fujikawa}, and $\mathcal{A}_{\rm CP}$ is determined by the fit.  \\
Results are compatible with the known values%, with precision dominated by sample size 
~(Tab. \ref{tab:res1}). In terms of yield per fb$^{-1}$ and peak purity, these results are comparable with Belle's best performance~\cite{Belle2013}, indicating a good understanding of early detector performance on charged-particle tracking, reconstruction of neutral pions, and flavor tagging.\\

%\clearpage

\section{CP violation in three-body decays}
\begin{figure}[h!]
 \centering
 \includegraphics[width=5.75cm]{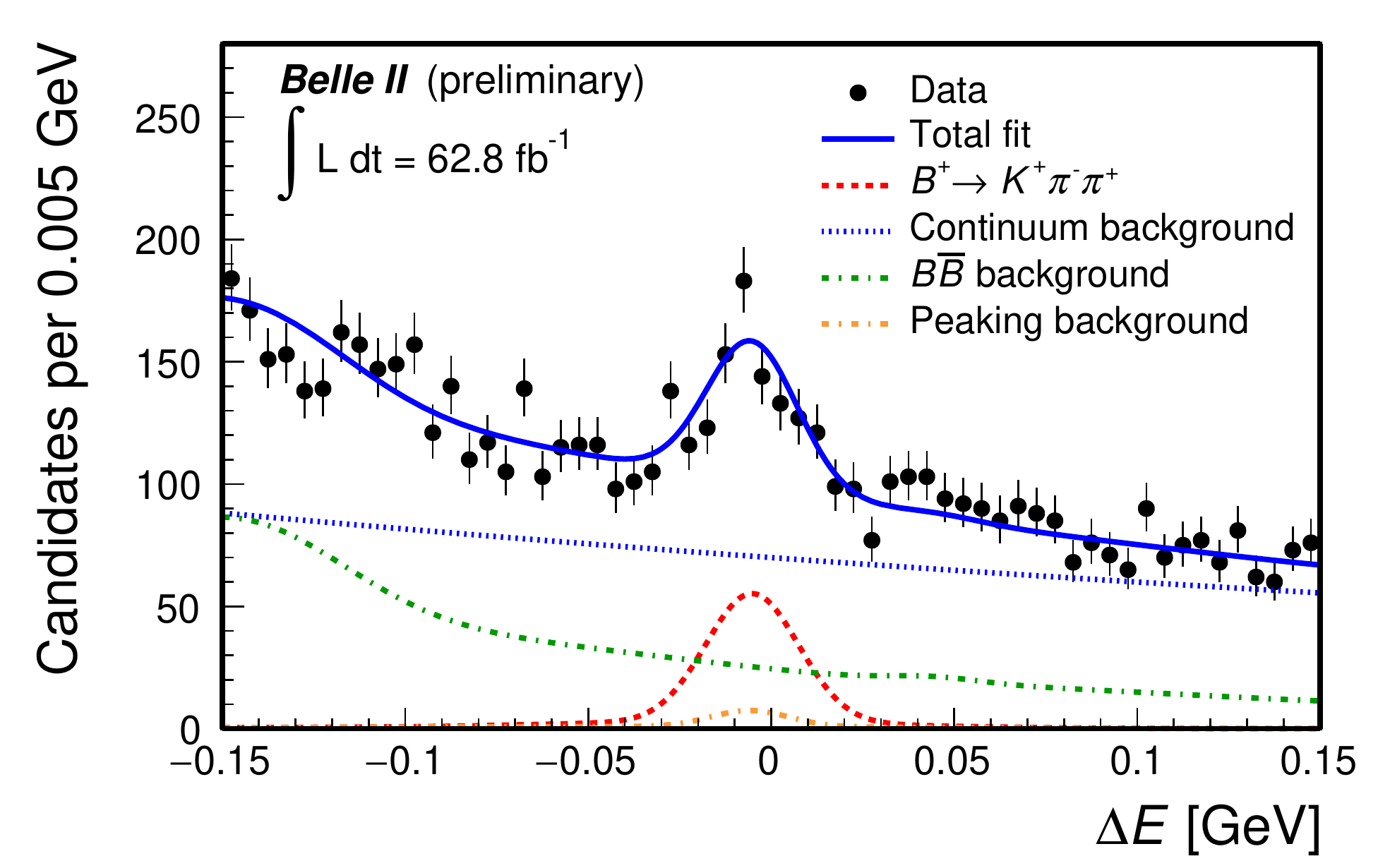} 
 \includegraphics[width=5.75cm]{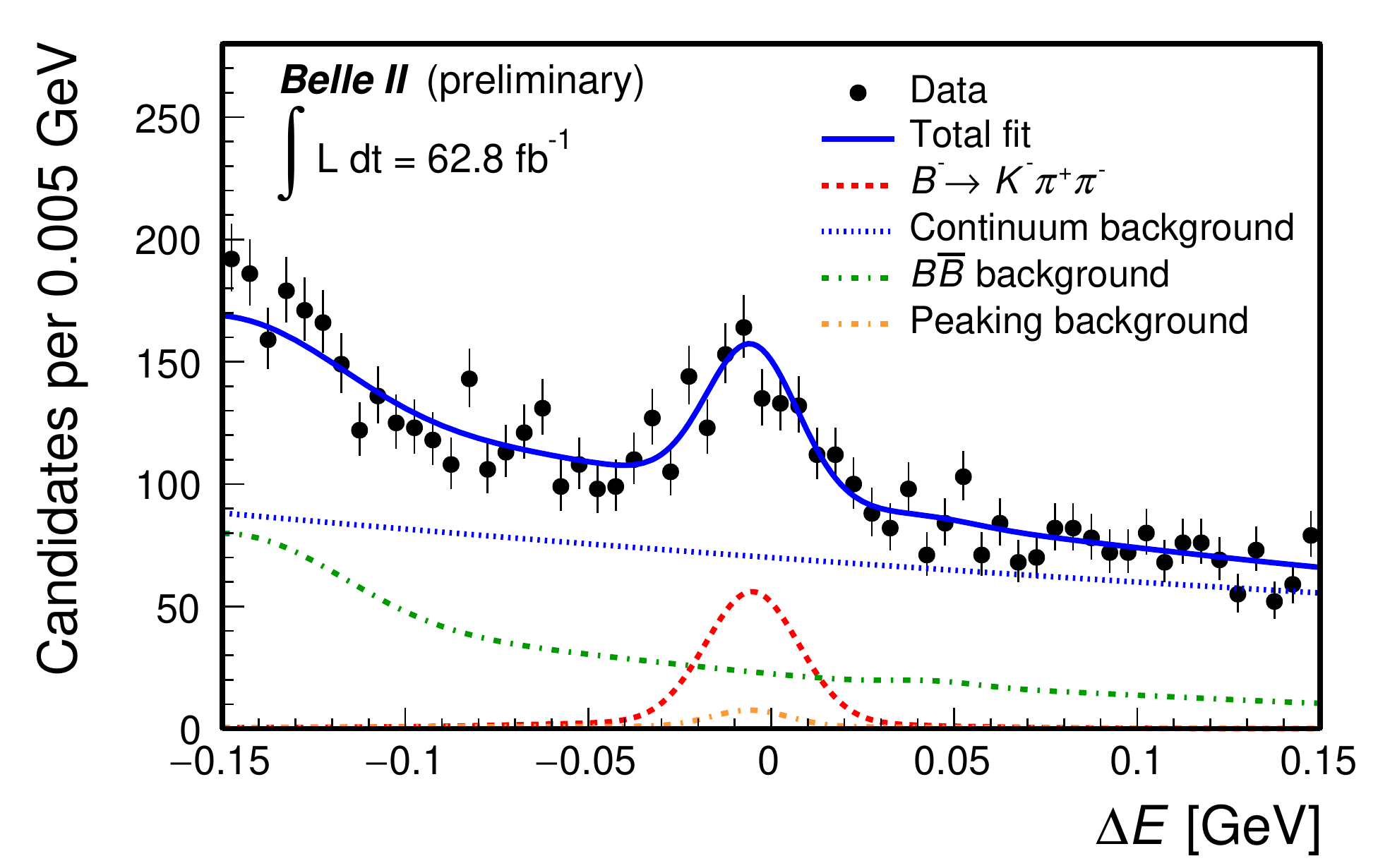}
 \includegraphics[width=5.75cm]{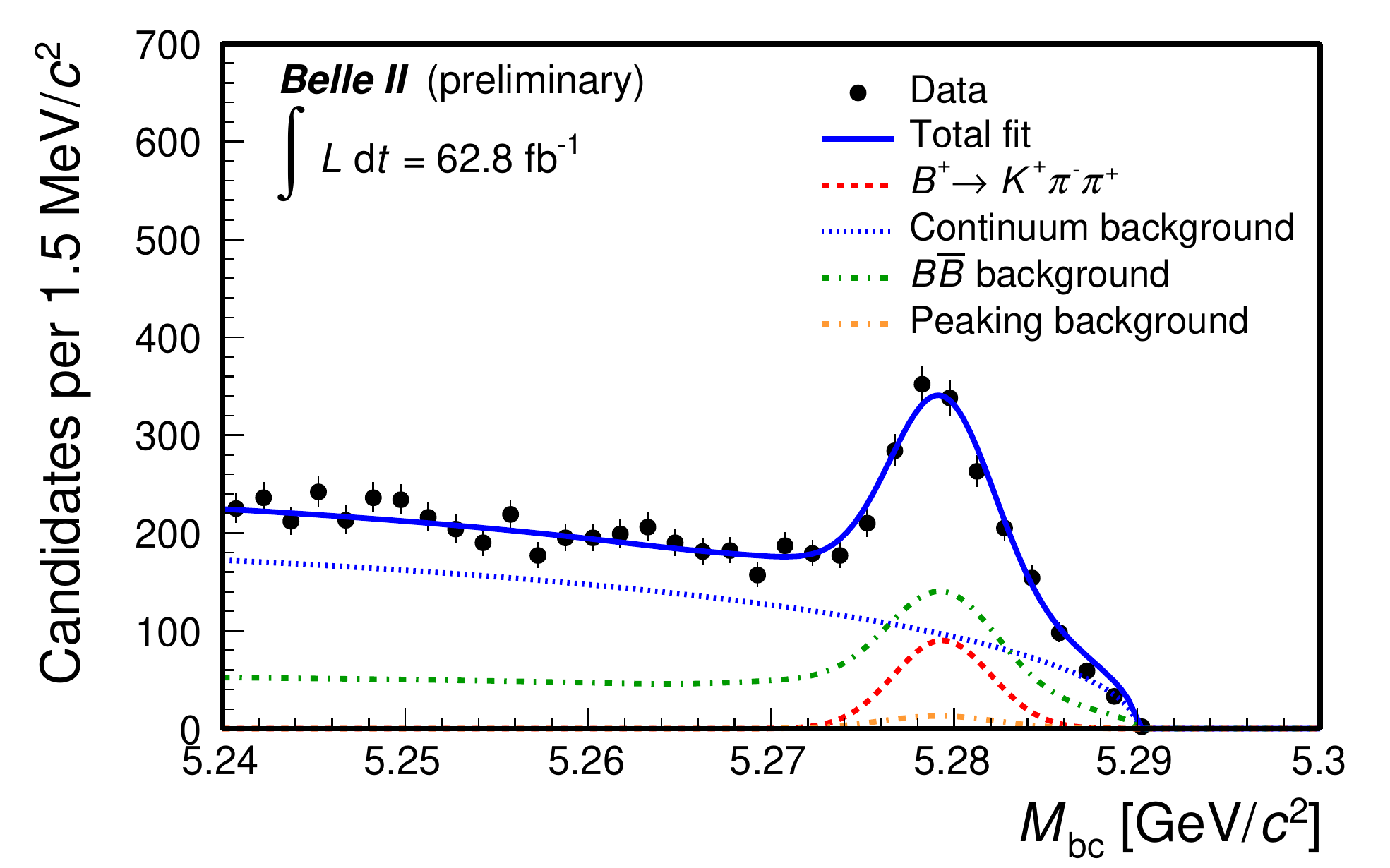} 
 \includegraphics[width=5.75cm]{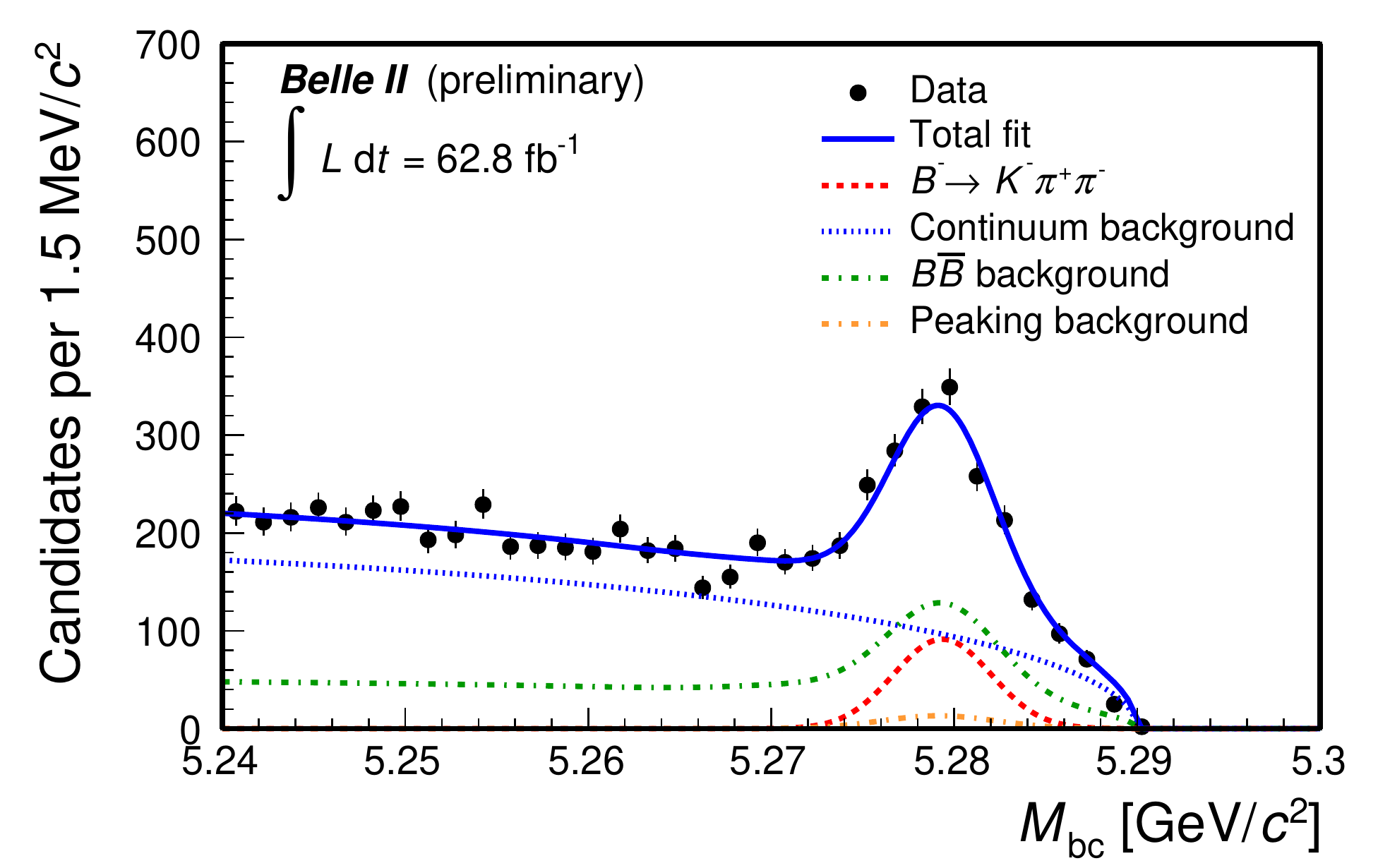} 
 \caption{Distributions of (top) $\Delta E$ and (bottom) $M_{\rm bc}$ for (left) $B^+\to K^+\pi^-\pi^+$ and (right) $B^-\to K^-\pi^+\pi^-$ candidates reconstructed in the 2019--2020 Belle~II data. Fit projections are overlaid.}
 \label{fig:Kpipi}
\end{figure}

Hadronic $B$ decays into multibody final state proceed through various charmless intermediate states, leading to manifestations of CP violation that vary over phase space. 
Measurements of localized CP asymmetries allow for studying the contributions of different processes, and probe non-SM physics.
Measurements of $B^+\to K^+K^-K^+$ and $B^+\to K^+\pi^-\pi^+$ decays are performed on the full data set. A challenge specific to these channels is the presence of peaking backgrounds from charmed $B$ decays yielding the same final state. We use the simulation to suppress such contamination, and include in the fit any remainder. We exclude the two-body mass ranges corresponding to $D^0$, $\eta_c$, and $\chi_{c1}$ decays for $B^+\to K^+K^-K^+$ and $D^0$, $\eta_c$, $\chi_{c1}$, $J/\psi$, and $\psi(2S)$ decays for $B^+\to K^+\pi^-\pi^+$.\\
Signal yields are extracted with two-dimensional maximum likelihood fits of the unbinned $\Delta E$ and $M_{\rm bc}$ distributions, and then used to determine $\mathcal{B}$ and $\mathcal{A}_{\rm CP}$ (Fig. \ref{fig:Kpipi}). Results are compatible with the known values%with precision limited by statistical uncertainties 
~(Tab.~\ref{tab:res1}).  They paved the way for extending these studies to multibody decays involving neutral final-state particles.\\

\section{Towards an $\alpha$/$\phi_2$ determination}
\begin{figure}[hb]
 \centering
 \includegraphics[width=5.75cm]{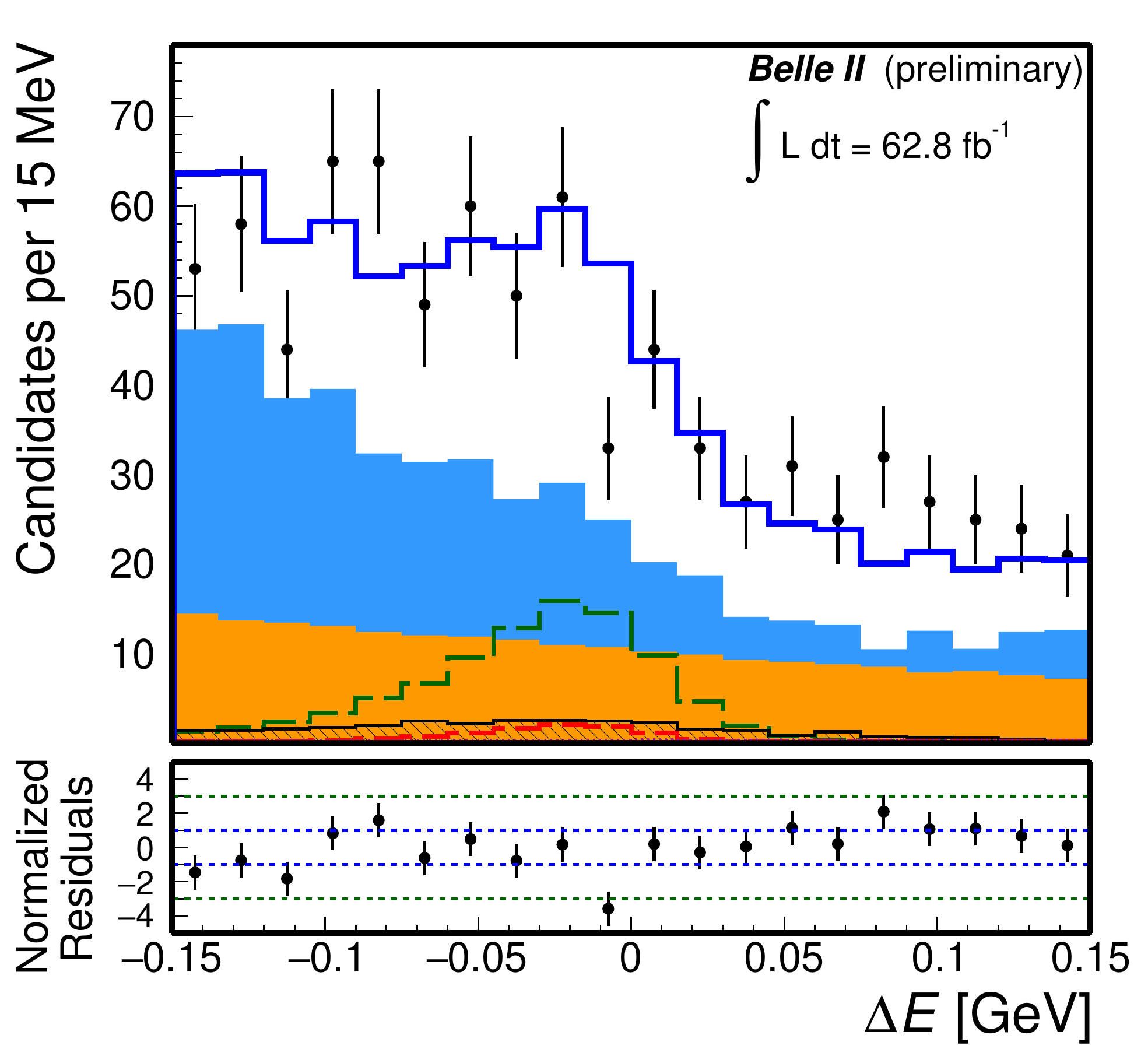} 
 \includegraphics[width=5.75cm]{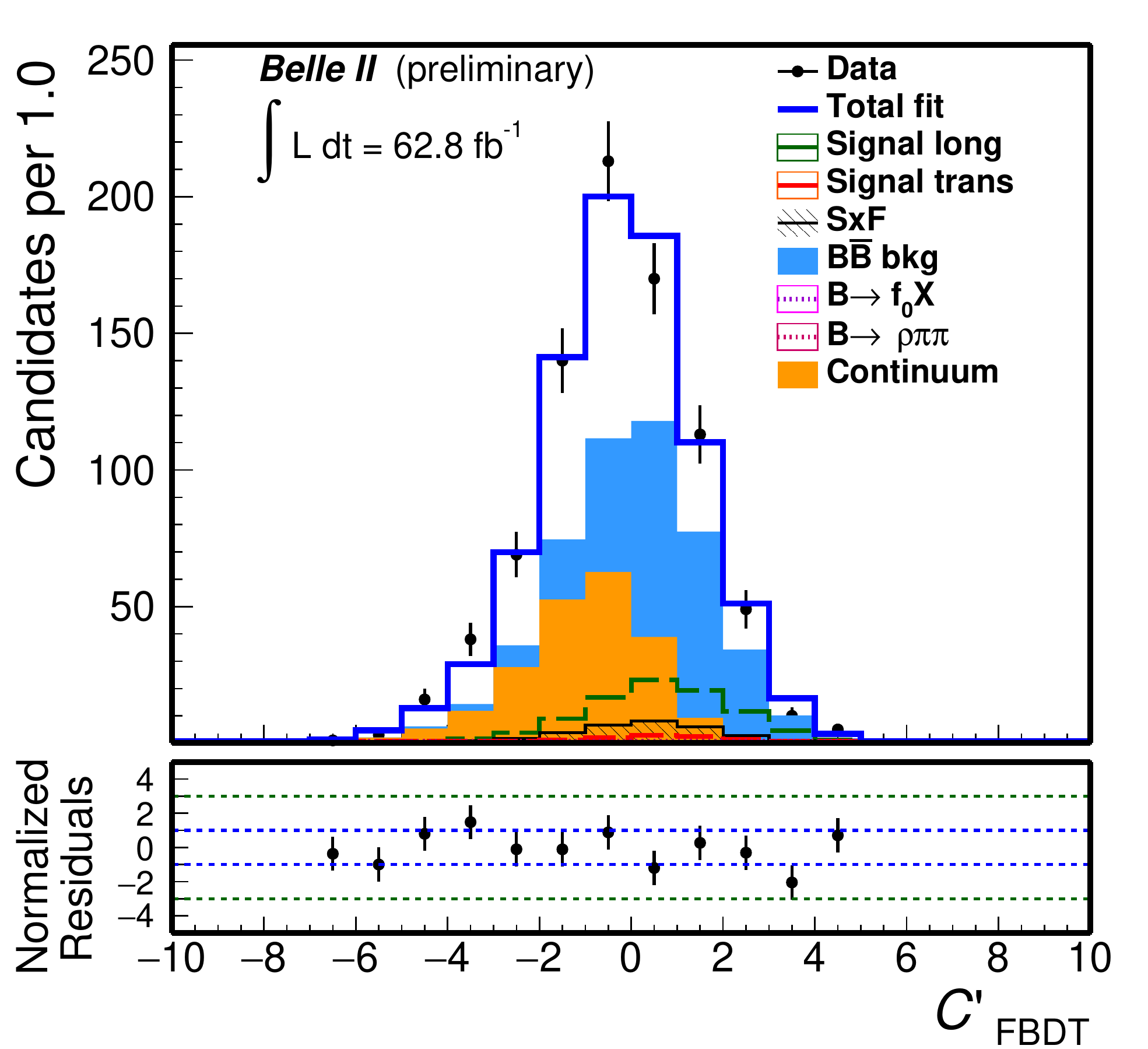} 
 \includegraphics[width=5.75cm]{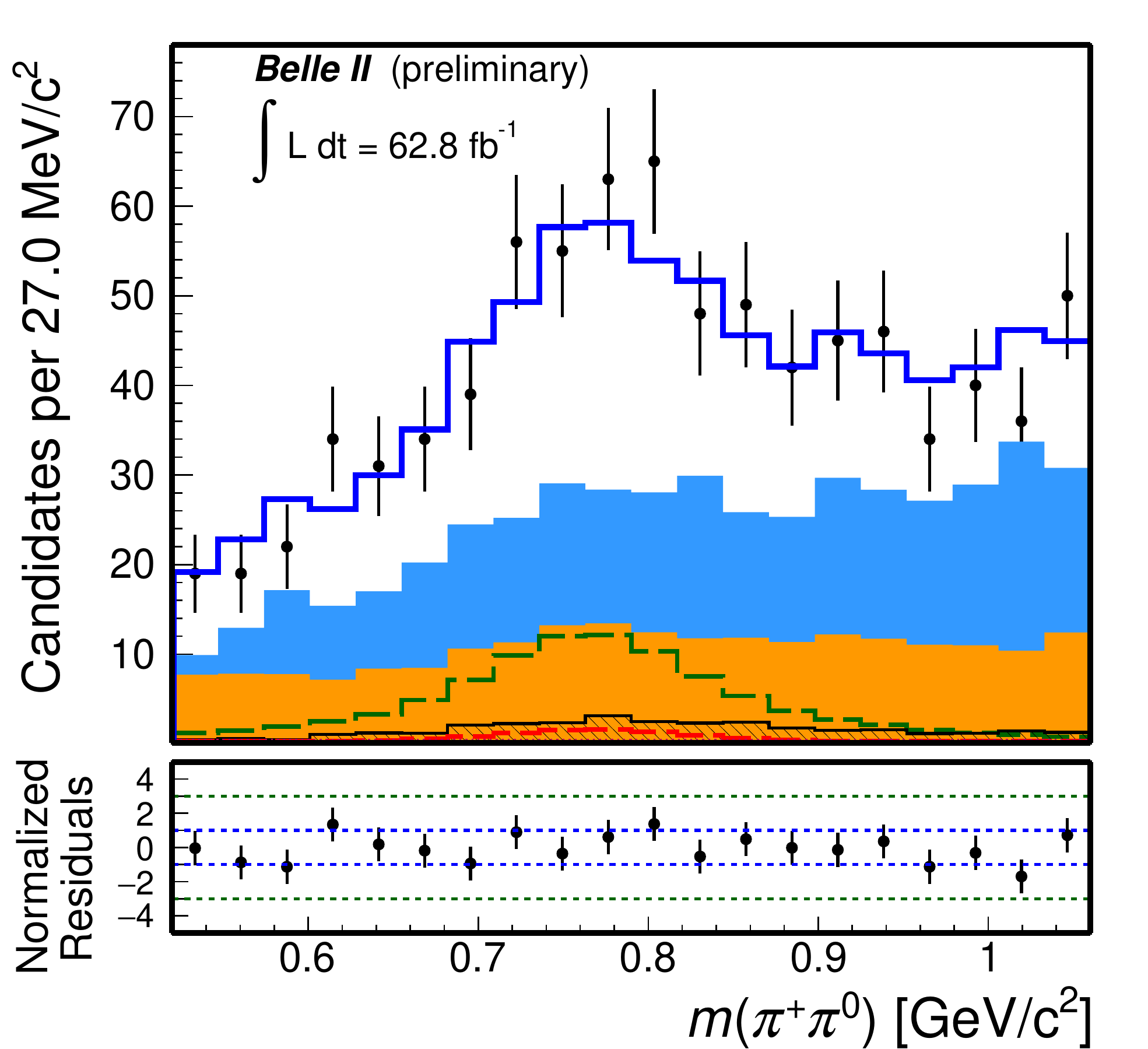}
 \includegraphics[width=5.75cm]{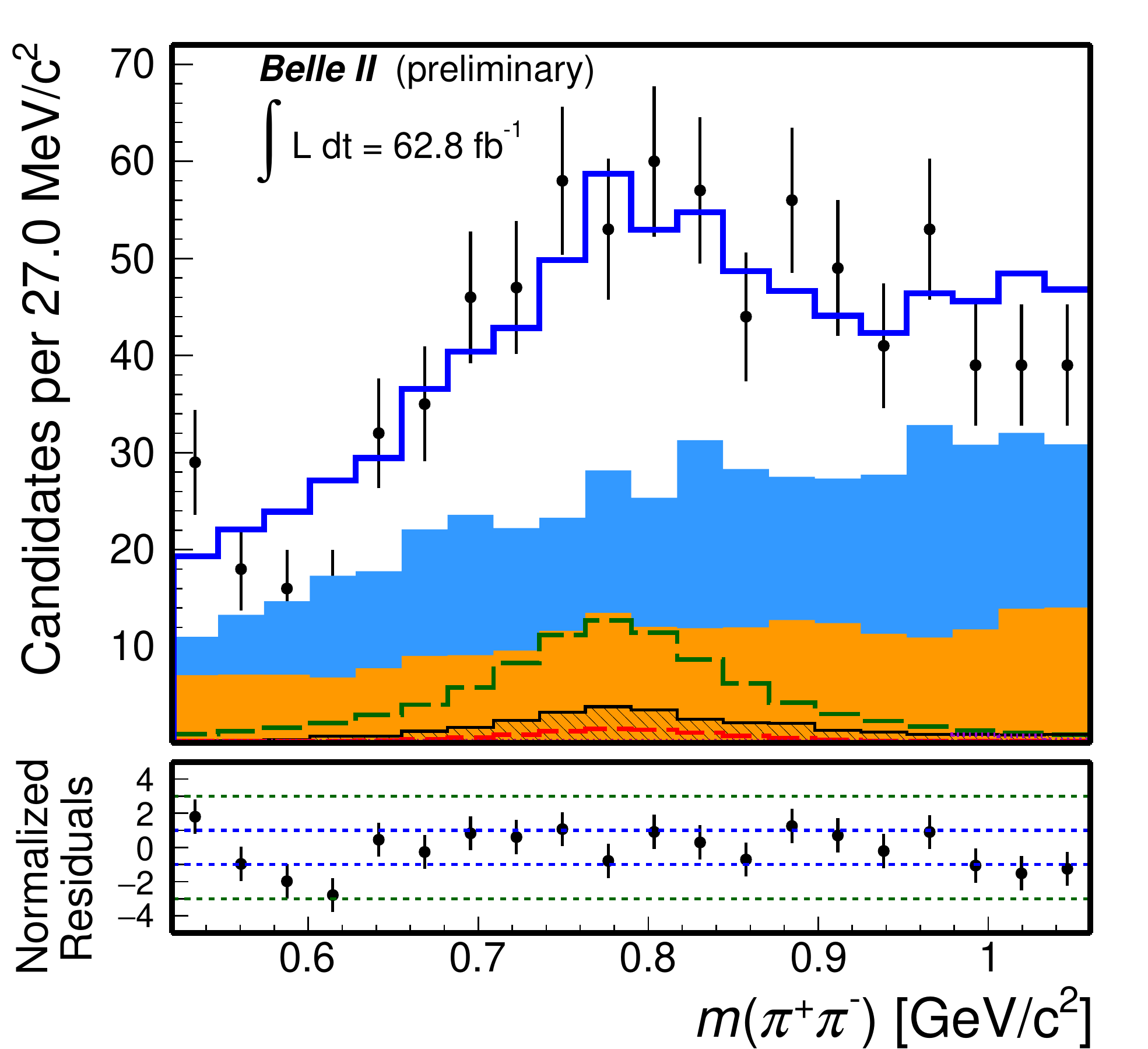}
 \includegraphics[width=5.75cm]{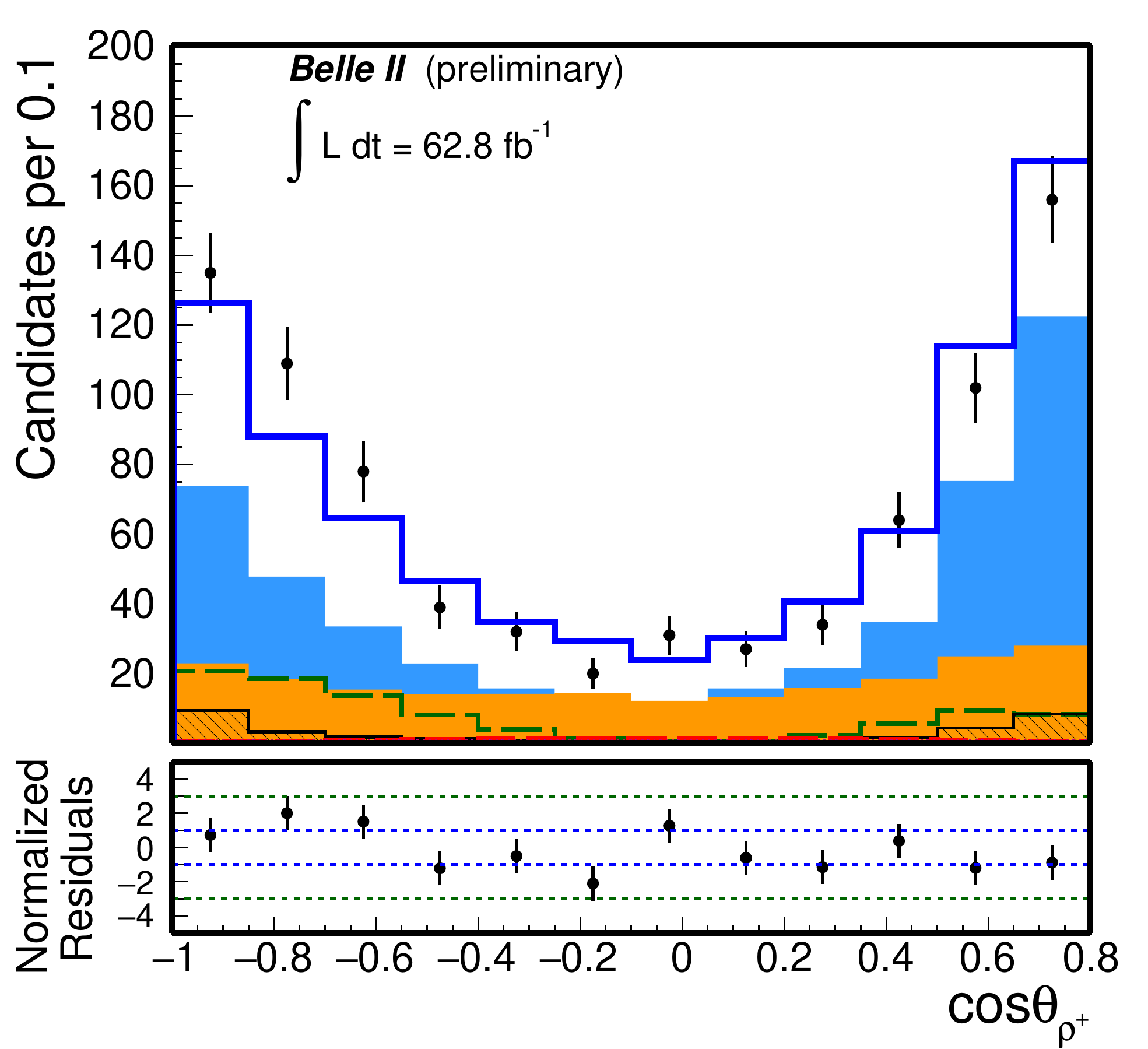} 
  \includegraphics[width=5.75cm]{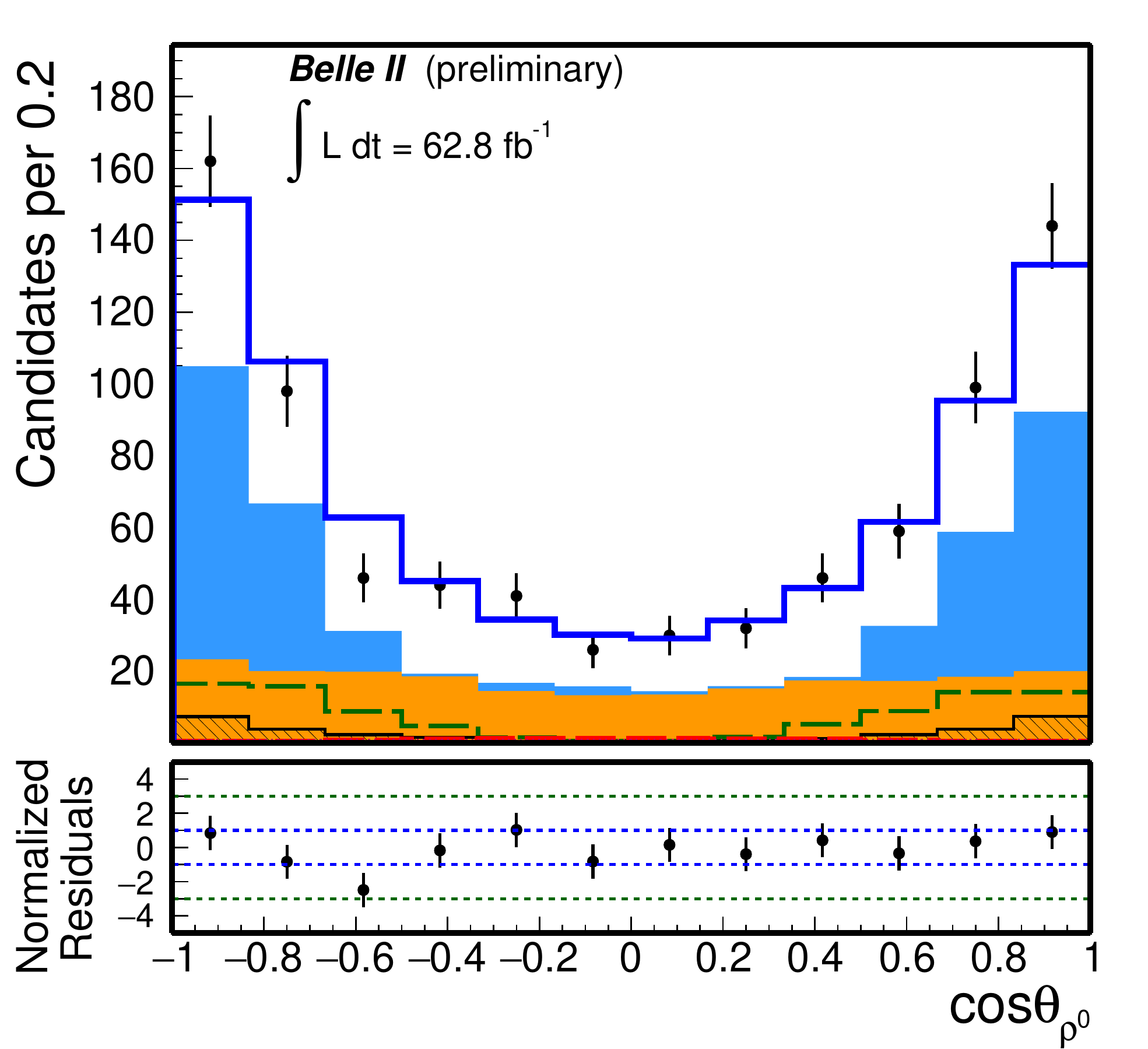}
 \caption{Distributions of (top left) $\Delta E$, (top right) continuum suppression output, (middle left) $m(\pi^+\pi^0)$, (middle right) $m(\pi^+\pi^-)$, (bottom left) $\cos\Theta_{\rho^+}$, (bottom right) $\cos\Theta_{\rho^0}$ for \mbox{$B^+ \to \rho^+\rho^0$} candidates reconstructed in the 2019--2020 Belle~II data. The self cross-feed component is indicated in the legend as ``SxF". Fit projections are overlaid.}
 \label{fig:rhorho}
\end{figure}

The CKM parameter $\alpha$/$\phi_2$ is associated with $b \to u\bar{u}d$ quark transitions, and is currently known with about $6\%$ precision~\cite{PDG}. Belle~II aims at determining $\alpha$/$\phi_2$ by measuring full sets of isospin-partner decays, such as $B\to \pi\pi$ and $B\to \rho\rho$. These decays are mediated by the same processes, but with subleading amplitudes of different relative intensities, offering attractive complementarity in measuring $\alpha$/$\phi_2$. 
In addition,  $B\to \rho\rho$ decays involve a spin-0 particle decaying into two spin-1 particles, resulting in three possible angular momentum configurations of the final state, which enrich the dynamics. We distinguish \textit{longitudinal} and \textit{transverse} polarization states, corresponding respectively to $\rho$ resonances having zero or opposite spin projections along the decay axis, and measure the fraction of longitudinally-polarized decays.\\
Results on $B^0\to \pi^+\pi^-$ and $B^+\to \pi^+\pi^0$ decays are obtained on the summer data set~\cite{charmless1}, while measurements of $B^+\to \rho^+\rho^-$ are based on the full data set. Reconstruction, selection, and fit for the $B\to\pi\pi$ measurements follow the general strategy. The measured branching fractions and CP asymmetries, shown in Table \ref{tab:res1}, are compatible with known values.% and dominated by the statistical uncertainties.  
~These results stringently benchmark the charged-particle-identification performances of the detector. \\
We reconstruct $B^+\to \rho^+\rho^-$ decays in the $\pi^+\pi^-\pi^+\pi^0$ final state. We form pion candidates applying baseline selection criteria, and combine them into $\rho$ candidates by requiring dipion-mass conditions. We exclude values close to 1 on the cosine of the helicity angle of the $\rho^+$ candidates, to suppress background from high-momentum charged particles combined with low-momentum $\pi^0$ candidates.  
We determine signal yields with a joint maximum likelihood fit to the unbinned distributions of $\Delta E$,  continuum-suppression output,   intermediate-resonance invariant masses $m(\pi^+\pi^0)$ and $m(\pi^+\pi^-)$, and cosines of the helicity angles of the intermediate resonances. The first four observables mainly separate signal from background; the other ones separate the two signal polarizations. The six-dimensional  probability density function is approximated as the product of one-dimensional densities. Sample components considered in the fit are transversely- and longitudinally-polarized signal, self cross-feed, background from $B\to f_0(980)X$ decays,  background from non-resonant $B\to\rho\pi\pi$ decays, remaining $B\bar{B}$~background, and continuum. 
We calculate the branching fraction $\mathcal{B} = \left(N_{\rm L}/\varepsilon_{\rm L} + N_{\rm T}/\varepsilon_{\rm T}\right)/\left(2 \cdot N_{B\bar{B}}\right)$ and the fraction of longitudinally polarized decays $f_{\rm L}=\left(N_{\rm L}/\varepsilon_{\rm L}\right)/\left(N_{\rm L}/\varepsilon_{\rm L} + N_{\rm T}/\varepsilon_{\rm T}\right)$, where $N_{\rm L}$ and $N_{\rm T}$ are longitudinally and transversely polarized signal yields, respectively, and $\varepsilon_{\rm L}$ and $\varepsilon_{\rm T}$ are the corresponding selection efficiencies determined from simulation. \\
The distributions of the six observables are shown in Fig. \ref{fig:rhorho} with fit projections overlaid. Results are compatible with known values (Tab. \ref{tab:res1}); they show significant improvements in precision with respect to the Belle measurement of $B^+\to \rho^+\rho^-$, based on a data sample of similar size~\cite{BelleRhorho}. These results validate both the early-phase detector performances and the capability of performing charmless analyses of increased complexity.\\

%\clearpage

\section{Summary}
We report on preliminary measurements of branching fractions, CP-violating charge asymmetries, and longitudinal polarization fractions in charmless $B$ decays at Belle~II. We use electron-positron collisions collected in 2019 and 2020 at the $\Upsilon(4S)$ resonance,  corresponding to integrated luminosities of up to 62.8 ${\rm fb^{-1}}$.  We extract signal yields for the decay modes $B^0\to K^+\pi^-$, $B^+\to K^+\pi^0$, $B^+\to K^{0}_{\rm S}\pi^+$, $B^0\to K^{0}_{\rm S}\pi^0$, $B^+\to K^+K^-K^+$, $B^+\to K^+\pi^-\pi^+$, $B^0\to \pi^+\pi^-$, $B^+\to \pi^+\pi^0$, and $B^+\to \rho^+\rho^0$. The results agree with known values and are dominated by statistical uncertainties. They show performance comparable or better than Belle's. This indicates a good understanding of the detector, and offers a reliable basis to assess projections for future performance.

\begin{table}[h]
\centering
\caption{Summary of results. The first contribution to the uncertainty is statistical, the second is systematic. The upper block refers to results on the summer data set (34.6 ${\rm fb^{-1}}$), the lower block to results on the full data set (62.8 ${\rm fb^{-1}}$).  For the $\mathcal{B}(B^+\to K^{0}\pi^+)$ and $\mathcal{B}(B^0\to K^{0}\pi^0)$ measurements, we consider a $0.5$ factor to account for the $K^0\to K^{0}_{\rm S}$ probability. }

\begin{tabular}{l c c c}
\hline\hline
 Decay mode & Branching fraction $\times~10^{-6}$ & CP-violating asymmetry & Longitudinal fraction \\
 \hline
 $B^{0}\to K^{+}\pi^{-}$ &  $18.9 \pm 1.4 \pm 1.0$ & $0.030 \pm 0.064 \pm 0.008$ & - \\[0.1cm]
 $B^+ \to K^+\pi^0$ & $12.7 ^{+2.2}_{-2.1}\pm 1.1$ & $0.052 ^{+0.121}_{-0.119}\pm 0.022$ & -  \\[0.1cm]
 $B^+ \to K^0\pi^+$ & $21.8 ^{+3.3}_{-3.0} \pm 2.9$ & $-0.072 ^{+0.109}_{-0.114} \pm 0.024$ & -  \\[0.1cm]
  $B^0 \to \pi^+\pi^-$ & $5.6 ^{+1.0}_{-0.9} \pm 0.3$ & -  & -  \\[0.1cm]
 $B^+ \to \pi^+\pi^0$ & $5.7 \pm 2.3\pm 0.5$ & $-0.268 ^{+0.249}_{-0.322}\pm 0.123$ & -  \\[0.1cm]
 \hline
 $B^0 \to K^0\pi^0$ & $8.5^{+1.7}_{-1.6} \pm 1.2$ & $-0.40^{+0.46}_{-0.44} \pm 0.04$  & -  \\[0.1cm]
 $B^+ \to K^+K^-K^+$  & $35.8 \pm 1.6 \pm 1.4$ & $-0.103 \pm 0.042 \pm 0.020 $ & -  \\[0.1cm]
 $B^+ \to K^+\pi^-\pi^+$ & $67.0 \pm 3.3\pm 2.3$ & $-0.010 \pm 0.050 \pm 0.021$ & -  \\[0.1cm]
 $B^+\to\rho^+ \rho^0$ & $20.6 \pm 3.2 \pm 4.0$& - & $0.936^{+0.049}_{-0.041} \pm 0.021$  \\[0.1cm]
\hline\hline
\end{tabular} 

\label{tab:res1}
\end{table}
\clearpage

%\subsection{Mathematics}
%Here is a lettered array~(\ref{e.all}), with eqs.~(\ref{e.house})
%and~(\ref{e.phi}):
%\begin{eqnletter}
 %\label{e.all}
 %\drm x_\sy{F} & = & 1.2\cdot10^3\un{cm}, \qquad
%                     \tx{where\ } \sy{F} = \tx{Fermi}    \label{e.house}\\
% \phi_i        & = & i\pi                                \label{e.phi}
%\end{eqnletter}

%\subsection{Citations}
%We're almost done, just some citations~\cite{ref:apo}
%and we will be over~\cite{ref:pul,ref:bra}.

%\acknowledgments
%The author acknowledge XXX, YYY.

\end{document}